\documentclass[pre,aps,twocolumn,superscriptaddress,longbibliography]{revtex4-1}
\usepackage[dvips]{graphicx}
\usepackage{amssymb,amsfonts,amsmath}
\usepackage{color}
\usepackage{ulem}
\usepackage{siunitx}
\usepackage[english]{babel}
\usepackage[hidelinks]{hyperref}

\newcommand{\nablarot}{\boldsymbol{\hat R}}
\newcommand{\Vext}{V_{\rm ext}}
\newcommand{\ui}{\hat{\textbf{u}}_1}
\newcommand{\uk}{\hat{\textbf{u}}_2}

\newcommand{\omegai}{\boldsymbol\omega_i}
\newcommand{\average}[1]{\left\langle\sum\limits_i #1 \right\rangle}
\newcommand{\ru}[1]{r_u^{#1}}
\newcommand{\vel}{\textbf{v}}
\newcommand{\deltar}{\delta(\textbf{r}-\textbf{r}_i)}
\newcommand\identity{1\kern-0.25em\text{l}}

\newcommand{\rv}{{\mathbf r}}
\newcommand{\uv}{{\hat{\mathbf u}}}
\newcommand{\deltau}{\delta(\uv-\uv_i)}
\newcommand{\tb}{{\mathbf t}}

\newcommand{\Jv}{{\bf J}}

\newcommand{\Fv}{{\bf F}}
\newcommand{\Tv}{{\bf T}}
\newcommand{\fv}{{\bf f}}
\newcommand{\ev}{{\bf e}}

\begin{document}

\title{Reduced-variance orientational distribution functions from torque sampling}	

\author{Johannes Renner}
\affiliation{Theoretische Physik II, Physikalisches Institut,
  Universit{\"a}t Bayreuth, D-95440 Bayreuth, Germany}

\author{Matthias Schmidt}
\affiliation{Theoretische Physik II, Physikalisches Institut,
  Universit{\"a}t Bayreuth, D-95440 Bayreuth, Germany}

\author{Daniel de las Heras}
\email{delasheras.daniel@gmail.com}
\homepage{www.danieldelasheras.com}
\affiliation{Theoretische Physik II, Physikalisches Institut,
  Universit{\"a}t Bayreuth, D-95440 Bayreuth, Germany}

\date{\today}

\begin{abstract}
We introduce a method to sample the orientational distribution function in computer simulations.
The method is based on the exact torque balance equation for classical many-body systems of interacting anisotropic particles in equilibrium.
Instead of the traditional counting of events, we reconstruct the orientational distribution function via an orientational integral of the torque acting on the particles.
We test the torque sampling method in two- and three-dimensions, using both Langevin dynamics and overdamped Brownian dynamics, and with two interparticle interaction potentials.
In all cases the torque sampling method produces profiles of the orientational distribution function with better accuracy than those obtained with the traditional counting method.
The accuracy of the torque sampling method is independent of the bin size, and hence it is possible to resolve the orientational distribution function with arbitrarily small angular resolutions.
\end{abstract}

\maketitle

\section{Introduction}
The spatial and orientational order in classical equilibrium many-body systems is the result of a delicate balance between forces and torques of internal, entropic (diffusive), and external origin.
One-body distribution functions, obtained as statistical averages resolved in either space, orientation or both of these, are essential for the  description and understanding of the organization of many-body systems at the microscopic level.
For example, the density profile, which is an average over a statistical ensemble of the number of particles at a given position, provides information about the spatial structure of the many-body system.
Traditionally, the density profile in computer simulations has been obtained by discretizing the simulation box and counting the number of particles in each element of the grid.
Since the structure of the many-body system is the result of a force balance, an alternative to counting events in order to obtain the density profile consists of reconstructing it from the spatially resolved force contributions~\cite{Borgis2013,Heras2018a}.
The density profiles obtained via force-sampling methods have a reduced variance as compared to those obtained via the traditional counting method.
Moreover, the density at a given position is constructed with information from the whole system.
As a result the error in the density profile does not depend on the size of the elements of the grid~\cite{Borgis2013,Heras2018a}.
The density profile can therefore be resolved with arbitrarily high spatial resolution without increasing the computational cost.
This is particularly useful for sampling two-~\cite{Heras2018a} and three-dimensional~\cite{Coles2019} density profiles.
Force-based estimators can be also used to improve the sampling of the radial distribution function~\cite{Borgis2013,Sutherland2021,Simon2022}, and that of the correlation functions required in the Green-Kubo expressions relevant for mobility profiles~\cite{Mangaud2020}. 

It is interesting to note that force-sampling methods can be derived from the general and versatile mapped averaging framework~\cite{Moustafa2015,Schultz2016,Schultz2019,Purohit2019,Moustafa2022}, in which approximate theoretical results are used to reformulate an ensemble average with reduced variance.
Reduced-variance estimators were first introduced in classical and quantum Monte Carlo simulations~\cite{Assaraf1999,Assaraf2007}.
An account of reduced-variance estimators that make use of force sampling methods is given in a recent review~\cite{Rotenberg2020}. 

Beyond constructing statistical estimators with low variance in equilibrium systems, the internal force can be used to derive force-based density functional theories~\cite{Tschopp2022,Sammueller2022}, and it plays a fundamental
role in the construction of exact sum rules using the symmetries of the system~\cite{Hermann2021,Hermann2022,Hermann2022a}. Moreover, the use of the thermodynamic force can also improve the accuracy of adaptive resolution schemes~\cite{Krekeler2018} in which the simulation box is split in regions that can be treated with different levels of resolution~\cite{Praprotnik2005,Potestio2013a,Potestio2013}. Another potential application of force sampling methods is to improve the convergence of Kirkwood-Buff~\cite{Kirkwood1951} integrals in molecular simulations~\cite{Dawass2020}.

Moreover, the knowledge of the internal force field is not only beneficial in equilibrium systems.
The adiabatic approximation, which substitutes the non-equilibrium internal forces by those in an equilibrium system, is at the core of popular dynamical theories such as dynamic density functional theory (DDFT)~\cite{evans1979,Marconi1999,Archer2004,Espanol2009,Vrugt2020}. 
Sampling the internal forces in many-body non-equilibrium simulations and comparing them to those in equilibrium systems is therefore crucial to develop and test the accuracy of dynamical theories that go beyond the adiabatic approximation such as superadiabatic-DDFT~\cite{Tschopp2022a} and power functional~\cite{Schmidt2013,Heras2020a,RevModPhys.94.015007} theories.
Knowledge of the non-equilibrium internal forces facilitates also the construction of the external force field that generates a desired dynamical response via custom flow methods~\cite{Heras2019,Renner2021}, and serves to gain insight into physical processes such as the occurrence of viscous forces generated by the acceleration field~\cite{Renner2022}.

In systems with translational and rotational degrees of freedom, such as liquid crystals, it is not only the forces but also the torques that are crucial in the determination of the equilibrium and non-equilibrium properties of the many-body system.
The force balance equation is complemented and coupled with a torque balance equation.
Together, the force and the torque balance equations determine in equilibrium the positional and the orientational order of the system.

Here, we demonstrate that torque sampling, i.e. the analogue to force sampling in systems with orientational degrees of freedom, significantly improves the sampling of the orientational distribution function in computer simulations as compared to traditional counting methods. 
As a proof of concept, we sample the torques using several differing types of dynamics (overdamped Brownian and Langevin dynamics), dimensionality (two- and three-dimensional systems), interparticle interaction potential (rectangular and Gay-Berne particles), type of orientational order (uniaxial and tetratic), and overall density. In all cases, torque sampling outperforms the traditional counting method.

\section{Theory}\label{sec:theory}

We consider here classical systems of $N$ identical interacting particles governed by either Langevin or overdamped Brownian dynamics.
Exact one-body force and torque balance equations hold in equilibrium, and can be used to calculate one-body distribution functions from the forces and torques acting in the system.
We start by revisiting the force balance equation in a many-body system with only translational degrees of freedom.

\subsection{Force balance equation for isotropic particles}
In many-body systems with only translational degrees of freedom, such as a system of isotropic particles (e.g.\ a fluid of Lennard-Jones particles), the exact one-body force density balance equation in equilibrium reads~\cite{Hansen2013}
\begin{equation}
	0 = -k_BT\nabla\rho(\rv) + \Fv(\rv).
  \label{EQfdeiso}
\end{equation}
The first term on the right hand side of Eq.~\eqref{EQfdeiso} stems from the (ideal gas) diffusion, with $k_B$ being the Boltzmann constant, $T$ is absolute temperature, $\nabla$ is the derivative with respect to the spatial coordinate $\rv$, and $\rho(\rv)$ is the one-body density distribution which is given by
\begin{equation}
  \rho(\rv) =
  \Big\langle \sum_i \delta(\rv-\rv_i)  \Big\rangle,
  \label{EQdensityDefinition}
\end{equation}
where the angles denote a statistical average over an equilibrium ensemble, $\delta(\cdot)$ is the Dirac distribution, $\rv_i$ is the position of particle $i$, and the sum runs over all the particles in the system.

The second term on the right hand side of Eq.~\eqref{EQfdeiso} is the force density profile, given by
\begin{equation}
 \Fv(\rv) = \Big\langle
  \sum_i\delta(\rv-\rv_i)\fv_i(\rv^N)
  \Big\rangle,
  \label{EQinternalForceDensityAsAverage}
\end{equation}
where $\fv_i$ is the sum of the internal and the external forces acting on particle $i$ in microstate $\rv^N=\rv_1\dots\rv_N$ with $N$ particles. That is
\begin{equation}
	\fv_i(\rv^N) = - \nabla_i u(\rv^N) + \fv_{\rm ext}(\rv_i),
\end{equation}
where $\nabla_i$ is the derivative with respect to $\rv_i$, and $u(\rv^N)$ is the total interparticle interaction potential.
In equilibrium, the (imposed) external force $\fv_{\rm ext}(\rv)$ must be conservative and hence 
\begin{equation}
	\fv_{\rm ext}(\rv) = -\nabla V_{\rm ext}(\rv),
\end{equation}
with $V_{\rm ext}(\rv)$ an imposed external potential.
The force profile follows directly from the force density profile via normalization with the density profile, i.e. $\fv(\rv)=\Fv(\rv)/\rho(\rv)$.

The sum of the ideal, internal, and external force densities vanishes everywhere in space since the system is in equilibrium.
Otherwise there would be a net flow of particles.
In equilibrium, the exact force density balance equation, Eq.~\eqref{EQfdeiso}, holds in systems following either Newtonian dynamics, Langevin dynamics, or overdamped Brownian dynamics.

Both, the density profile $\rho(\rv)$ and the force density profile $\Fv(\rv)$, can be easily sampled in computer simulations via Eqs.~\eqref{EQdensityDefinition} and~\eqref{EQinternalForceDensityAsAverage}, respectively.
Sampling $\rho(\rv)$ via Eq.~\eqref{EQdensityDefinition} is the traditional method of counting of events of particle occurrences at space points.

The exact force density balance equation, Eq.~\eqref{EQfdeiso}, can also be used to calculate the density profile $\rho(\rv)$ via the forces instead of the direct traditional counting method.
Inverting Eq.~\eqref{EQfdeiso} results in 
\begin{equation}
	\rho(\rv)=\rho_0 + (k_BT)^{-1}\nabla^{-1}\cdot\Fv(\rv),
	\label{EQfs}
\end{equation}
with $\rho_0$ a constant and $\nabla^{-1}$ the inverse $\nabla$ operator.
In effectively one-dimensional systems (e.g.\ planar geometry), the profiles depend only on one space coordinate and hence the $\nabla^{-1}$ operator reduces to a simple spatial integral.
Different approaches can be used to solve Eq.~\eqref{EQfs} in more general geometries~\cite{Heras2018a}.
The unknown integration constant $\rho_0$ in Eq.~\eqref{EQfs} can be determined via normalization of the density
\begin{equation}
\int d\rv\rho(\rv)=N,
\end{equation}
where the integral is over the whole system volume.
Results for the density profile calculated via force sampling, Eq.~\eqref{EQfs}, carry a statistical uncertainty smaller than that of the standard counting method~\cite{Heras2018a} since (i) force sampling avoids the inherent ideal gas fluctuations, and
(ii) uses non-local information, the forces in the whole system, to determine the density profile at each space point.

\subsection{Torque balance equation for anisotropic particles}
For anisotropic particles, the one-body density distribution depends not only on the space coordinates $\rv$ but also on the orientation, which is denoted here by the unit vector $\uv$:
\begin{equation}
	\rho(\rv,\uv)=\left\langle\sum_i\deltar\deltau\right\rangle.\label{eq:rhoru}
\end{equation}
In addition to the exact equilibrium one-body force density balance equation,
\begin{align}
	0=-k_BT\nabla\rho(\rv,\uv)+\Fv(\rv,\uv),\label{eq:forcebalance}
\end{align}
there exists an exact one-body torque density balance equation:
\begin{align}
	0=-k_BT\nablarot\rho(\rv,\uv)+\Tv(\rv,\uv).\label{eq:torquebalance}
\end{align}
Here, $\Fv(\rv,\uv)$ and $\Tv(\rv,\uv)$ are the force density and the torque density, respectively. Both, $\Fv(\rv,\uv)$ and $\Tv(\rv,\uv)$, contain external and internal (inter-particle) contributions and they depend in general on position and orientation. As before, $\nabla$ is the gradient operator acting on the position, and $\nablarot$ is the orientational counterpart operator acting on the orientation $\uv$, i.e.
\begin{eqnarray}
	\nablarot&=&\uv\times\nabla_\uv\label{eq:R},
\end{eqnarray}
with $\nabla_\uv$ the derivative with respect to the Cartesian coordinates of $\uv$.

The one-body torque density is accessible in computer simulations via
\begin{align}
	\Tv(\rv,\uv)&=\left\langle\sum_{i}\deltar\deltau\tb_i(\rv^N,\uv^N)\right\rangle,\label{eq:Ttotalav}
\end{align}
with $\uv^N=\uv_1\dots\uv_N$ and $\uv_i=(\sin\theta_i\cos\varphi_i,\sin\theta_i\sin\varphi_i,\cos\theta_i)$ being the orientation of particle $i$. Here, $\theta_i$ and $\varphi_i$ are the polar and azimuthal angles of particle $i$, respectively. The torque on particle $i$ is $\tb_i$, given by
\begin{align}
    \tb_i\left(\rv^N,\uv^N\right)&=-\nablarot_i u\left(\rv^N,\uv^N\right)-\nablarot_i\Vext\left(\rv_i,\uv_i\right),
\end{align}
with $\nablarot_i=\uv_i\times\nabla_{\uv_i}$.
Note that both the total interparticle potential $u(\rv^N,\uv^N)$ and the external potential $\Vext(\rv,\uv)$ are allowed to carry a dependence on the particle orientation.
The one-body torque density is therefore the sum of internal and external contributions
\begin{equation}
	\Tv(\rv,\uv)=\Tv_{\text{int}}(\rv,\uv)+\Tv_{\text{ext}}(\rv,\uv),
\end{equation}
with 
\begin{align}
	\Tv_{\text{int}}(\rv,\uv)&=-\left\langle\sum_{i}\deltar\deltau\nablarot_iu\left(\rv^N,\uv^N\right)\right\rangle,\\
	\Tv_{\text{ext}}(\rv,\uv)&=-\left\langle\sum_{i}\deltar\deltau\nablarot_i\Vext\left(\rv_i,\uv_i\right)\right\rangle.
\end{align}
Using eq.~\eqref{eq:rhoru} the external contribution is simply
\begin{equation}
	\Tv_{\text{ext}}(\rv,\uv)=-\rho(\rv,\uv)\nablarot\Vext(\rv,\uv).\label{eq:Text}
\end{equation}

Further details regarding the derivation of the one-body torque density balance in equilibrium are given in Appendix~\ref{B:tbe}.

In general, the force and the torque density balance equations are linked via the one-body density distribution. 
Here, we focus only on the role of the torque balance equation.
For this we consider in what follows systems that are homogeneous in space and therefore cases in which the force balance equation does not play any role.
In such systems $\rho(\rv,\uv)=\rho_bf(\uv)$, with $\rho_b$ being the bulk density, and $f(\uv)$ being the orientational distribution function.
That is, $f(\uv)d\uv$ is the probability of finding a particle with orientation $\uv$ within a solid angle $d\uv$.
The orientational distribution function is therefore normalized such that
\begin{equation}
\int d\uv f(\uv)=1.
\end{equation}
Using the traditional sampling method, the orientational distribution function can be sampled in computer simulations as
\begin{eqnarray}
	f(\uv) & = & \frac1N\left\langle\sum_{i}\delta\left({\uv}-{\uv_i}\right)\right\rangle\\
	&=&\frac{1}{N}\left\langle\sum_{i}\frac{1}{\sin\theta_i}\delta\left({\theta}-{\theta}_i\right)\delta\left({\varphi}-{\varphi}_i\right)\right\rangle.\label{eq:odf}
\end{eqnarray}
The prefactor $1/N$ ensures the proper normalization of the orientational distribution function.

For spatially homogeneous systems, the one-body torque density balance equation~\eqref{eq:torquebalance} simplifies to
\begin{equation}
   0=-k_BT\rho_b\nablarot f(\uv) + \textbf{T}(\uv).\label{eq:tbhomogeneous}
\end{equation}

Isolating the ideal gas term of Eq. (\ref{eq:tbhomogeneous}) and integrating appropriately we obtain an expression for the orientational distribution function
\begin{align}
    f(\uv)&=f_0+(\rho_bk_BT)^{-1}\nablarot^{-1}\cdot\textbf{T}(\uv).
\end{align}
Here, $\nablarot^{-1}$ is formally the inverse operator of $\nablarot$ and $f_0$ is an integration constant that ensures the proper normalization of the orientational distribution function.
For two-dimensional systems and uniaxial three-dimensional systems, the inverse operator  $\nablarot^{-1}$ reduces to a simple angular integral, see Appendix~\ref{A:interaction}.

Obtaining the orientational distribution function via the one-body torque density balance has the advantage of treating the ideal gas part explicitly and hence, it avoids the corresponding fluctuations present in the counting method. The only source of statistical inaccuracies is in the sampled one-body torque density which is integrated over in order to obtain the orientational distribution function.
As it turns out, this process reduces the statistical noise significantly.

\begin{figure*}
    \centering
    \includegraphics[width=\linewidth]{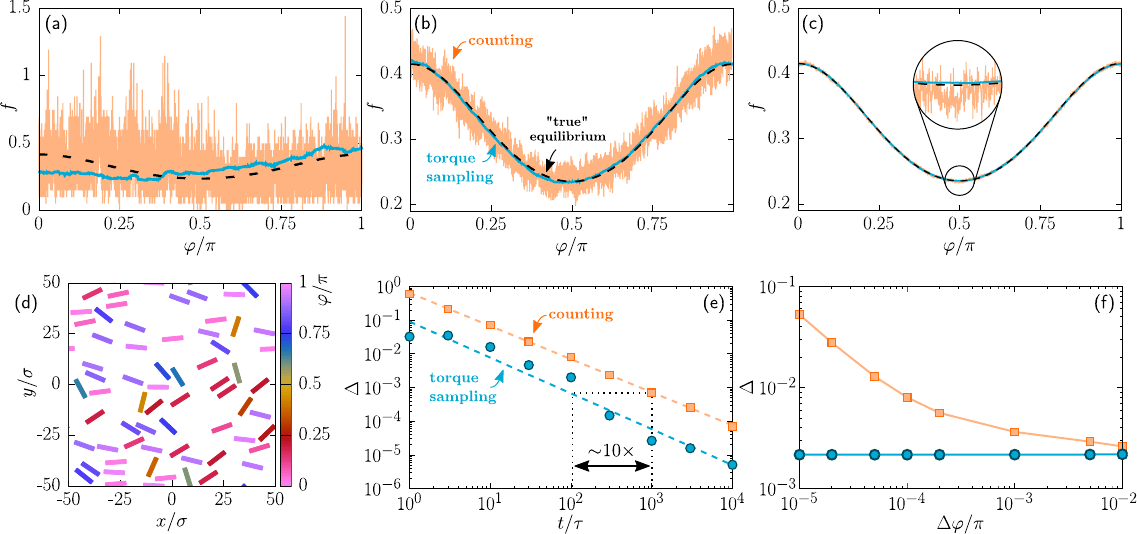}
	\caption{Orientational distribution function sampled with the counting method (orange) and the torque sampling method (blue) for three different sampling times: 
	(a) $10\tau$, (b) $10^3\tau$, and (c) $10^5\tau$. The black dashed line is the ''true'' equilibrium profile obtained by sampling over $10^7\tau$ and taking the arithmetic mean of the counting and the torque sampling methods. The bin size is $10^{-4}\pi$. The inset in (c) is a close view of the encircled region.
	(d) A characteristic snapshot of the system: $N=64$ particles with rectangular shape subject to a weak external potential that orients the particles along the $x-$axis.
	A $\log-\log$ plot of the error $\Delta$ as a function of the sampling time (e) and the error $\Delta$ as a function of the bin size (f) using the counting (orange squares) and the torque sampling (blue circles) method.
	In panel (e) the bin size is fixed to $10^{-4}\pi$ and the dashed lines are linear fits. In panel (f) the sampling time is fixed to $10^2\tau$ and the solid lines are guides for the eye.}
    \label{fig1}
\end{figure*}

\section{Results}\label{sec:Results}
As a proof of concept, we test the validity of the torque sampling method with two different systems: (i) two-dimensional rectangular particles following Langevin dynamics and (ii) three-dimensional Gay-Berne particles following overdamped Brownian dynamics. 

\subsection{Two-dimensional system of rectangular particles}
We consider a two-dimensional system of particles with rectangular shape undergoing Langevin dynamics (implemented according to Ref.~\cite{GroenbechJensen2013}).
The interaction between two particles is modeled via a purely repulsive potential $\phi(r)=\epsilon\left(\frac{\sigma}{r}\right)^{12}$. Here, $r$ is the minimum distance between the two particles, $\sigma$ is
our length scale, and $\epsilon$ is our energy scale.
The potential acts only between the two closest points (one on each particle) located on the particles' perimeter.
The interparticle potential generates both an internal force and an internal torque.
Details about the calculation of the forces and the torques, as well as about the integration of the equations of motion are given in the Appendix~\ref{C:interaction}

We study a system of $N=64$ rectangular particles with length $L/\sigma=10$ and width $D/\sigma=2$ in a square box of length $100\sigma$ and periodic boundary conditions.
We set the temperature to $k_BT/\epsilon=1$ and the integration time step to $\Delta t/\tau=10^{-3}$ with $\tau=\sigma\sqrt{m/\epsilon}$ and $m$ the mass of one particle.
We sample every $10\,\Delta t$.
Since the system is very diluted, the equilibrium bulk state is isotropic.
We induce orientational order via the external potential $\Vext(\varphi)/\epsilon=-0.5\cos^2\varphi$, with the angle $\varphi$ measured anticlockwise with respect to the $x-$axis.
A characteristic snapshot of the system is shown in Fig. \ref{fig1}.

\begin{figure*}
    \centering
    \includegraphics[width=\linewidth]{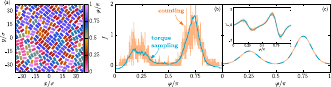}
	\caption{(a) Characteristic snapshot of a Langevin dynamics simulation of $N=290$ rectangular particles in a square box of side length $L_\text{box}/\sigma=75$ subject to an external potential
	that favors particle orientations along the bottom-right to top-left diagonal of the box.
	The particles are colored according to their orientation $\varphi$, measured with respect to the $x-$direction, see colorbar.
	Orientational distribution $f(\varphi)$ obtained via counting (orange lines) and torque sampling (blue lines) using a bin size of $10^{-3}\pi$ and for two sampling times: $1\tau$ (b) and $10^5\tau$ (c).
	The inset in (c) shows the numerical angular derivative of the orientational distribution function $f'(\varphi)=df/d\varphi$ using the central difference.}
    \label{fig2}
\end{figure*}

We initialize the particles randomly and equilibrate the system with a simulation lasting $10^3\tau$.
After equilibration we sample the orientational distribution function via the counting and the torque sampling methods.
The results are shown in Fig.~\ref{fig1} for three different sampling times: $10\tau$ panel (a), $10^3\tau$ panel (b), and $10^5\tau$ panel (c).
Due to the head-tail symmetry of the particles we represent the orientational distribution function in the interval $\varphi\in[0,\pi]$ only.
Torque sampling provides at each time a profile which is closer to the ''true'' equilibrium profile than the one provided by the counting method.
The ''true'' equilibrium profile $f_{\rm eq}(\varphi)$ is defined here as the arithmetic mean of the profiles obtained with the counting and the torque sampling methods in a long simulation (total simulation time $10^7\tau$).
For all sampling times the statistical noise in the profiles using the counting method is significantly larger than that using the torque sampling method.

To quantify the accuracy of each method, we define an error parameter as the integrated square difference between the ''true'' equilibrium profile and the sampled profile 
\begin{align}
	\Delta=\int_0^{\pi} d\varphi\left[f_s(\varphi)-f_\text{eq}(\varphi)\right]^2.
\end{align}
Here, $f_s$ is the profile sampled using the counting or the torque sampling methods.
As can be seen in Fig. \ref{fig1}(e) the error of the torque sampling method is for all sampling times below the error of the counting method.
For this particular bin size ($10^{-4}\pi$) one has to sample about $10$ times longer using the counting method than using the torque sampling method to reach the same accuracy.

In Fig.~\ref{fig1}(f) we investigate the effect of varying the bin size at a fixed sampling time ($10^2\tau$).
By decreasing the bin size we increase the level of detail with which we resolve the orientational distribution function.
However, decreasing the bin size obviously increases the number of bins and, as a direct consequence, the error in the traditional counting method also increases.
Note that in the counting method the number of events that contribute to each bin is proportional to the bin size.
On the other hand, the error in the torque sampling method is essentially independent of the bin size.
The error does not increase by decreasing the bin size because the orientational distribution function is not determined by the local number of events.
Instead, at each orientation the orientational distribution function is obtained via an orientational integral over the torque density.
Analogue behaviour occurs also when sampling the density profile using the force sampling method in systems with only translational degrees of freedom~\cite{Borgis2013,Heras2018a,Rotenberg2020}.

{\bf Tetratic order.}
Instead of sampling the complete, angle-resolved, orientational distribution function, it is common to sample only a reduced set of orientational order parameters (moments of the distribution).
However, having access to the complete orientational distribution function can help to fully understand the type of order in the system.
To illustrate this, we investigate a densely packed system of $N=290$ particles with length $L/\sigma=4$ and width $D/\sigma=2$ in a square box of length $75\sigma$.
The equilibration time was $10^4\tau$.
Due to their small length-to-width aspect ratio, the particles form in bulk at moderate densities a tetratic phase~\cite{MartinezRaton2005,MartinezRaton2009,GonzalezPinto2019}.
In the tetratic phase the particles are equally likely oriented along two directions perpendicular to each other.
We add an external potential of the form
$\Vext(\varphi)/\epsilon = -0.5\sin^2(\varphi-\varphi_0)$ with $\varphi_0/\pi=1/4$ and set the temperature to $k_BT/\epsilon=1$. 
The external potential breaks the symmetry of the tetratic phase by favoring the orientation along the bottom-right to top-left diagonal of the square simulation box.

A snapshot of the system is shown in Fig.~\ref{fig2}(a). The particles are colored according to their orientation.
The resulting orientational distribution function is shown in Fig. \ref{fig2}(b) for a short sampling time of $1\tau$ and in Fig.~\ref{fig2}(c) for a sampling time of $10^3\tau$.
Clearly more particles are aligned along the bottom-right to top-left diagonal ($\varphi/\pi=0.75$) than along the other diagonal ($\varphi/\pi=0.25$) due to the external potential.
In this example, the uniaxial order parameter or even the combination of both the uniaxial and the tetratic order parameters would not give enough information about the orientational order in the system.

The distributions sampled with torque sampling are always smoother than those sampled with the counting method.
However, torque sampling sometimes produces artifacts for very short sampling times (of the order of $1\tau$), like the negative values around $\varphi/\pi=0$ shown in Fig.~\ref{fig2}(b).
It might be possible to eliminate these artifacts by either using a combination of linear estimators~\cite{Coles2021} or the mapped averaging framework~\cite{Purohit2019}.
The artifacts are at least partially due to local angular currents orginated by fluctuations that do not vanish (on average) due to the short sampling times.
The occurrence of these angular currents is apparent when comparing the orientational distribution functions sampled at short, Fig.~\ref{fig2}(b), and long, Fig.~\ref{fig2}(c), sampling times (Cf.\ the evolution of the value of the orientational distribution functions at the peaks).
For longer sampling times, Fig.~\ref{fig2}(c), the angular current averages to zero for all orientations, and the distribution function calculated with torque sampling is free of artifacts.
The profile obtained with torque sampling is more precise than that obtained via counting.
Even at very long sampling times, e.g.\ $10^5\tau$ in~\ref{fig2}(c), torque sampling outperforms counting. 
This is particularly clear when looking at the numerical angular derivative of the distribution function, see inset of Fig.~\ref{fig2}(c).

\begin{figure}
    \centering
    \includegraphics[width=0.9\linewidth]{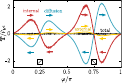}
	\caption{Components of the torque balance equation (normalized with the bulk density $\rho_b$) as a function of the angle in the tetratic configuration with an external field shown in Fig.~\ref{fig2}. The torques point in the $z$-direction. Positive (negative) torques try to rotate the particles anticlockwise (clockwise), as indicated at selected angles by the color arrows. The external potential favors particle alignments along the bottom right  to top left diagonal ($\varphi/\pi=0.75$) of the simulation box. The bottom left to top right diagonal is located at $\varphi/\pi=0.25$. Shown are the internal torque density (red), the diffusive torque density (blue), the external torque density (yellow), and the total torque density (black).}
    \label{fig3}
\end{figure}

Sampling the torques is not only useful to improve the sampling of the orientational distribution function but it also helps to understand the underlying physics. 
As an illustration, we show in Fig.~\ref{fig3} the components of the torque balance equation in the system with tetratic ordering and an external potential.
The torques point along the $z-$direction. That is, positive (negative) torques tend to rotate the particles anticlockwise (clockwise), increasing (decreasing) therefore the value of $\varphi$.
The diffusive torque (blue) always favors an isotropic state by trying to remove the inhomogeneities in the orientational distribution function.
In the current configuration, the diffusive torque tries to orient the particles away from the diagonals.
The behaviour of the internal torque depends on several factors such as the interparticle potential, the temperature, and the density.
In the current example, the internal torque (red) favors tetratic ordering by trying to align the particles along the diagonals.
The imposed external torque (yellow) tries to orient the particles along the bottom-right to top-left diagonal ($\varphi/\pi=0.75$) and it also tries to orient the particles away from the other diagonal at $\varphi/\pi=0.25$.
As dictated by the torque balance equation, the sum of all three components (diffusive, internal, and external) vanishes since the system is in equilibrium, see Fig.~\ref{fig3} (black line).

\subsection{Three-dimensional Gay-Berne fluid}
We further test the method in a three-dimensional system of $N=500$ Gay-Berne particles~\cite{Gay1981} confined in a box of size lengths $L_x/\sigma_0=10$, $L_y/\sigma_0=10$, and $L_z/\sigma_0=25$ with periodic boundary conditions.
We use the parameters $\sigma_0$ and $\epsilon_0$ of the Gay-Berne potential as our length and energy scales, respectively.
All details about the interparticle potential are presented in Appendix~\ref{A:Gay-Berne}.
We set the length-to-width ratio of the particles to three.
The particles follow overdamped Brownian dynamics.
Time is measured in units of $\tau_0=\gamma\sigma_0^2/\epsilon_0$, with $\gamma$ the translational friction coefficient against the implicit solvent. 
The particles are subject to an external potential $\Vext(\theta)/\epsilon_0=-0.5\cos^2(\theta)$, with $\theta$ the polar angle.
Hence, the external potential favors uniaxial alignment of the particles along the $z$-axis.
The temperature is set to $k_BT/\epsilon_0=0.5$.
For details regarding the implementation of the overdamped Brownian dynamics see Appendix~\ref{A:BD}

\begin{figure*}
    \centering
    \includegraphics[width=\linewidth]{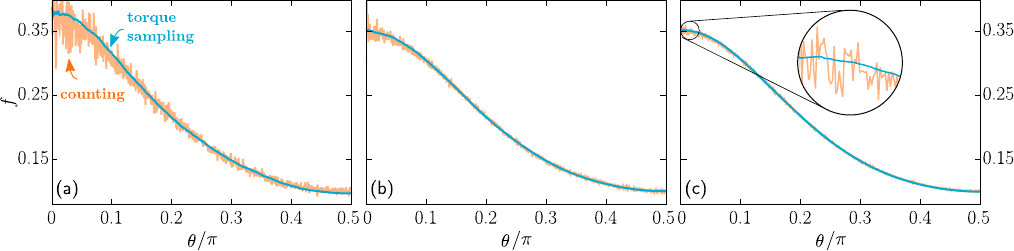}
    \caption{Overdamped Brownian dynamics simulation of $N=500$ Gay-Berne particles in a three-dimensional box with periodic boundary conditions.
	Orientational distribution function $f$ as a function of the polar angle $\theta$ obtained via counting (orange line) and torque sampling (blue line) using a bin size of $\Delta\theta=10^{-3}\pi/2$ for different sampling times: $10^{3}\tau_0$ (a), $10^4\tau_0$ (b), and $10^5\tau_0$ (c).The inset in panel (c) is a close view of the region of small polar angles as indicated.}
    \label{fig4}
\end{figure*}

The orientational distribution functions obtained via torque sampling and counting are shown for different sampling times in Fig.~\ref{fig4}.
Again, torque sampling provides profiles with better accuracy than counting.
The differences between both methods are more acute for small values of the polar angle.
The area of the bins on the unit sphere decreases close to the poles.
Therefore, less events contribute to each bin, which produces large fluctuations of the profile obtained with the counting method.
However, the profile obtained with torque sampling is unaffected by this problem since the error is independent of the bin size.

\section{Conclusion}
Reduced-variance estimators can be constructed using force sampling methods~\cite{Rotenberg2020} to measure e.g.\ the density profile and the radial distribution function in computer simulations with better accuracy than the traditional counting method.
We have shown here that in equilibrium systems of interacting anisotropic particles, reduced-variance estimators can be also constructed via torque sampling.
By sampling the torques and using the exact torque balance equation of equilibrium many-body systems, we have developed a method to accurately reconstruct the orientational distribution function.
Although the cases that we have studied here are arguably toy models, they do cover a wide range of situations, including two- vs three-dimensional systems, dilute vs dense systems, uniaxial vs tetratic orientational order, and Langevin vs overdamped Brownian dynamics. In all cases, torque sampling has outperformed counting.

Force sampling works equally well in Brownian dynamics, molecular dynamics, and Monte Carlo simulations~\cite{Heras2018a}.
Hence, although we have used here Brownian and Langevin dynamics, the torque sampling method is expected to also outperform the counting method in molecular dynamics and Monte Carlo simulations.

For small bin sizes, the statistical error for the counting method diverges, while the error for the torque sampling method is independent of the size of the bin.
Hence, torque sampling can be particularly useful in cases where a large number of bins might be required such as for example when investigating biaxial nematics in three-dimensional systems~\cite{Allen1990,Berardi2008,Berardi2011}.  

We have restricted our study to cases without positional order such that force and torque balance equations are decoupled.
There exist several fully inhomogeneous standard situations accessible in computer simulations~\cite{Wilson2005,Care2005,Allen2019,Zannoni2022} in which both the density profile and the orientational distribution profile depend on the position coordinate, i.e. $\rho(\rv,\uv)=\rho(\rv)f(\rv,\uv)$.
These include, among others, the formation of stable bulk phases with both positional and orientational order~\cite{Veerman1990,McGrother1996,Chiappini2019}, confinement~\cite{Wall1997,Trukhina2008,Geigenfeind2015}, sedimentation~\cite{Savenko2004,Beek2004,ViverosMendez2014}, formation of topological defects~\cite{Dzubiella2000,Andrienko2001,Garlea2016,Monderkamp2021}, and nucleation~\cite{Takahashi2021} in liquid crystals. 
The force balance equation and the torque balance equation are then coupled and jointly determine the spatial and the orientational order of the system.
The combination of force and torque sampling should be in such cases substantially better than counting which requires filling a multidimensional histogram in both positions and orientations.

The formulation of the torque sampling method presented here cannot be directly applied to hard particle models~\cite{Mederos2014}, in which forces arise only due to particle collisions. 
However, the mapped averaging framework is applicable in hard particle models~\cite{Trokhymchuk2019}. 
Using the torque balance equation as input for the mapped averaging framework might result in reduced-variance estimators for the orientational distribution function.
Exact sum rules involving the torques follow from the symmetries of the system~\cite{Hermann2021} and might be also useful in the derivation of reduced-variance estimators in computer simulations of anisotropic particles.

The forces between individual colloidal particles are also accessible experimentally~\cite{Dong2022}.
It might therefore be possible to use force and torque sampling methods for the determination of distribution functions in experimental systems.

\section*{Data availability}
All the data supporting the findings are available from the corresponding author upon
reasonable request.

\begin{acknowledgments}
  We greatly acknowledge Enrique Velasco for helping us during the implementation of the Langevin dynamics simulation code, for suggesting the interaction potential between the rectangles, and for useful discussions. We also thank Daniel Borgis for useful discussions.
  This work is supported by the German Research Foundation (DFG) via project number 447925252.
\end{acknowledgments}

\appendix 
\section{Force density and torque density balance equations}\label{B:tbe}
The force balance and torque balance equations in the context of density functional theory are shown e.g.~\ in Ref.~\cite{Rex2007a}.
We sketch here the derivation of the force density and torque density balance equations for many-body systems of particles following overdamped Brownian dynamics.
Let $\Psi(\rv^N,\uv^N,t)$ be the many-body probability distribution function of an overdamped system. Then, statistical averages $\langle\cdot\rangle$ within the Fokker-Planck formalism can be computed as
\begin{equation}
	\langle\cdot\rangle = \int d\rv^N\int d\uv^N \cdot \Psi(\rv^N,\uv^N,t),\label{eq:avg}
\end{equation}
where the integrals cover the complete space of configurations.
The velocity $\vel_i$ and the angular velocity $\omegai$ of particle $i$ are given by
\begin{align}
	\gamma\vel_i &= -k_BT\nabla_i\ln\Psi-\nabla_iu(\rv^N,\uv^N)+\fv_{\rm ext}(\rv_i,\uv_i,t),\label{eq:vi}\\
	\gamma_r\omegai &= -k_BT\nablarot_i\ln\Psi-\nablarot_iu(\rv^N,\uv^N)+\boldsymbol{\tau}_{\rm ext}(\rv_i,\uv_i,t).\label{eq:wi}
\end{align}
Here, $\gamma$ and $\gamma_r$ are the translational and rotational friction constants against the implicit solvent, respectively,
and $\boldsymbol{\tau}_{\rm ext}$ is an external torque field.
The current $\Jv$ and the angular current $\Jv_{\omega}$ are
\begin{align}
	\Jv(\rv,\uv,t)&=\average{\vel_i\deltar\deltau},\\
	\Jv_{\omega}(\rv,\uv,t)&=\average{\omegai\deltar\deltau}.
\end{align}
Multiplying equations~\eqref{eq:vi} and~\eqref{eq:wi} by $\deltar\deltau$, summing over all particles $i$, and applying the average in Eq.~\eqref{eq:avg} yields directly
\begin{align}
	\gamma\Jv(\rv,\uv,t)&=-k_BT\nabla\rho(\rv,\uv,t)+\Fv(\rv,\uv,t),\label{eq:J}\\
	\gamma_r\Jv_{\omega}(\rv,\uv,t)&=-k_BT\nablarot\rho(\rv,\uv,t)+\Tv(\rv,\uv,t).\label{eq:Jo}
\end{align}
The above force~\eqref{eq:J} and torque~\eqref{eq:Jo} density balance equations hold in full non-equilibrium situations.
In equilibrium, the equations simplify further since: (i) the time dependence drops from the density profile, the force density $\Fv$, and the torque density $\Tv$,
(ii) the external force and the external torque are conservative, and (iii) both $\Jv$ and $\Jv_{\omega}$ vanish.
Hence, in equilibrium Eqs.~\eqref{eq:J} and~\eqref{eq:Jo} reduce to Eqs.~\eqref{eq:forcebalance} and~\eqref{eq:torquebalance}, respectively.

The equilibrium force and torque balance equations do not change if the particles obey Langevin or molecular dynamics instead of Brownian dynamics
but the derivation is slightly different. 
To derive the force and the torque balance equations in Langevin dynamics or molecular dynamics, one needs to time derive the current and the angular current, both of which also vanish in equilibrium:
\begin{align}
	{\dot\Jv}(\rv,\uv)&=\frac{d}{dt}\average{\vel_i\deltar\deltau}=0,\label{eq:jMD}\\
	{\dot\Jv_{\omega}}(\rv,\uv)&=\frac{d}{dt}\average{\omegai\deltar\deltau}=0.\label{eq:jMDs}
\end{align}
Here, the average $\langle\cdot\rangle$ is again performed over the complete configuration space which in molecular dynamics includes integrals over the linear and angular momenta in addition to those
over the positions and the orientations of the particles.
Incorporating the time derivative inside the averages in Eqs.~\eqref{eq:jMD} and~\eqref{eq:jMDs} results in the force and torque balance equations.
In equilibrium, the integrals over the linear and the angular momenta can be carried out explicitly.

\section{Torque sampling for single angular dependencies}\label{A:interaction}
We derive here the expressions for the orientational distribution function as an angular integral over the torques in the system.
The rotational operator can be written as
\begin{eqnarray}
	\nablarot=\uv\times\nabla_\uv
	=\textbf{e}_\varphi\dfrac{\partial }{\partial \theta}-\textbf{e}_\theta\dfrac{1}{\sin\theta}\dfrac{\partial }{\partial \varphi}\label{eq:R2},
\end{eqnarray}
with $\textbf{e}_\varphi$ and $\textbf{e}_\theta$ being the unit vectors on the unit sphere in the azimuthal and in the polar directions, respectively.\\

\noindent{\bf Two-dimensional system.} In the two-dimensional system of rectangular particles, the orientational distribution
function depends only on the azimuthal angle $f=f(\varphi)$. Hence, using $\theta=\pi/2$, the rotational operator, eq.~\eqref{eq:R2}, simplifies to $\nablarot=\ev_z\partial/\partial\varphi$, and
the torque density balance equation~\eqref{eq:tbhomogeneous} is then
\begin{align}
	0 &= -k_BT\rho_b\ev_z\frac{\partial f(\varphi)}{\partial\varphi}+\Tv(\varphi),\label{eq:bla}
\end{align}
with $\Tv$ also directed along the $\ev_z$ direction.
The orientational distribution function can be then reconstructed with the sampled torques via
\begin{equation}
	f(\varphi)=f_0+\frac1{k_BT\rho_b}\int d\varphi \Tv(\varphi)\cdot\ev_z,
\end{equation}
with $f_0$ a normalization constant such that $\int d\varphi f(\varphi)=1$. 

Using Eq.~\eqref{eq:Text} it follows that
\begin{equation}
	\Tv_{\text{ext}}(\varphi)=-\rho_bf(\varphi)\frac{\partial\Vext(\varphi)}{\partial\varphi}\ev_z,
\end{equation}
which inserted in Eq.~\eqref{eq:bla} can be used to first solve the homogeneous equation analytically and then treat the internal torque density as an inhomogeneity.
We find that both approaches give similar results.\\

\noindent{\bf Three-dimensional uniaxial system.} We consider now the three dimensional system of Gay-Berne particles with a uniaxial distribution function, i.e. $f=f(\theta)$ and an external potential $\Vext(\theta)$ that depends also only on the polar angle. Hence the rotational operator, Eq.~\eqref{eq:R2}, is $\nablarot=\ev_\varphi\partial/\partial\theta$, which inserted in the torque density balance equation yields

\begin{equation}
	0 = -k_BT\rho_b\frac{\partial f(\theta)}{\partial\theta}\ev_\varphi + \Tv,\label{eq:bla3}
\end{equation}
where $\Tv=(T_{\text{int}}(\theta)+T_{\text{ext}}(\theta))\ev_\varphi$, from which we obtain
\begin{equation}
	f(\theta)=f_0+\frac{1}{k_BT\rho_b}\int d\theta T(\theta).
\end{equation}
The normalization constant $f_0$ is here such that $\int d\uv f(\theta)=1$.
Again, it is possible to first analytically solve the homogeneous differential equation of~\eqref{eq:bla3} by writing the external torques explicitly
\begin{equation}
	\Tv_{\text{ext}}(\theta)=-\rho_b f(\theta)\frac{\partial\Vext(\theta)}{\partial\theta}\ev_\varphi,
\end{equation}
and treat the internal part as the inhomogeneous part.

\section{Interparticle interaction between two rectangles and Langevin dynamics}\label{C:interaction}
\begin{figure}
    \centering
    \includegraphics[width=0.8\linewidth]{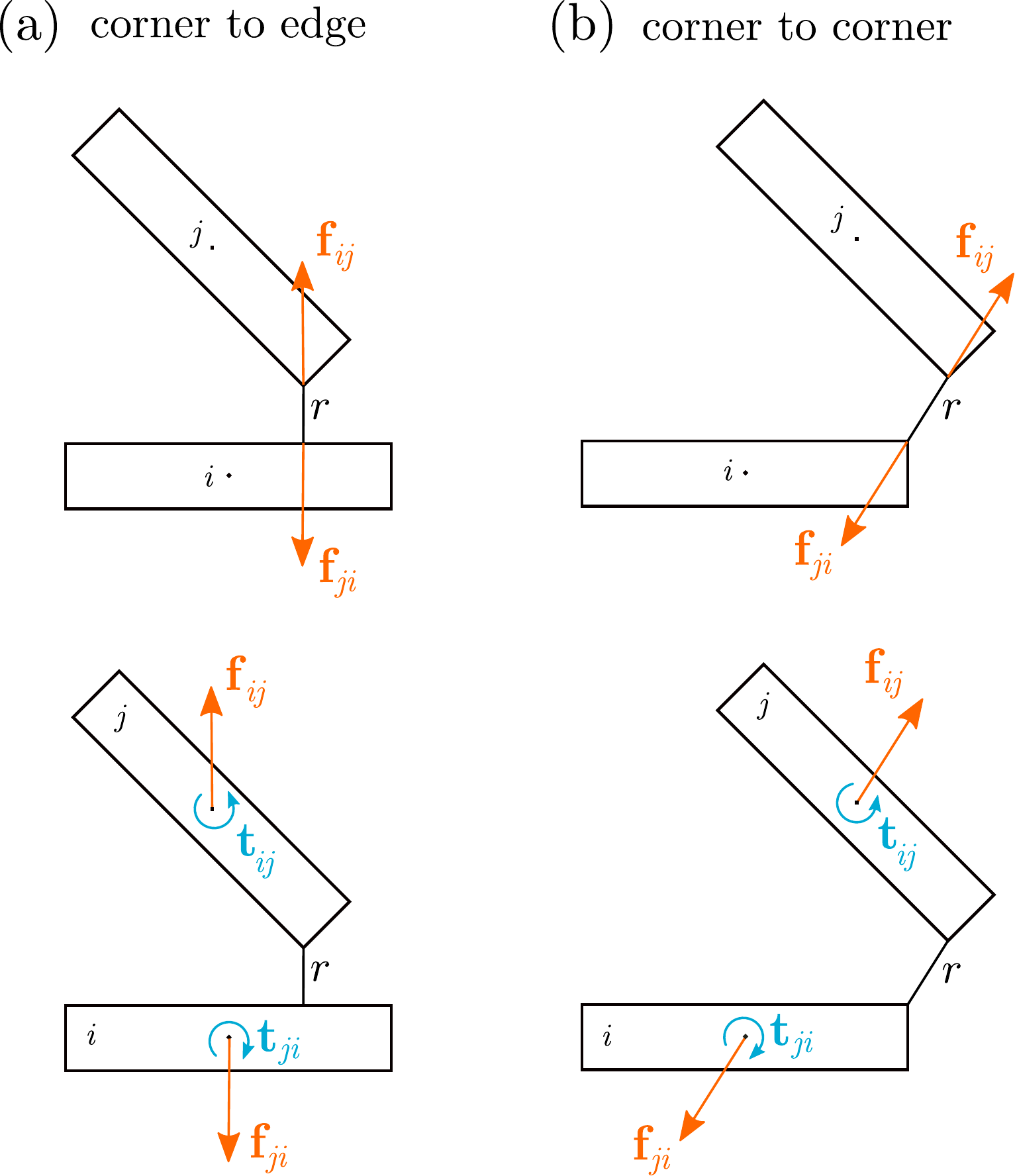}
	\caption{The minimum distance between two rectangles $r$ is either between a corner and an edge (a) or between two corners (b).
	The effect of the contact force acting on the points of minimum distance (top panels) is equivalent to the effect of the same force and a torque acting on the center of mass (bottom panels).}
    \label{fig5}
\end{figure}

Two rectangles interact via a purely repulsive pair-potential of the form 
\begin{align}
    \phi(r)=\epsilon\left(\frac{\sigma}{r}\right)^{12},
\end{align} with $r$ being the minimum distance between the two rectangles.
Depending on the positions and the orientations of the particles, there are two possible scenarios, see Fig.~\ref{fig5}: (i) the minimum distance is between a corner of one particle and a point located on an edge of the other particle, or (ii) the minimum distance is between two corners.
We introduce a cut-off distance of $r_c=2L+3\sigma$ between the centers of mass of the particles.
The potential generates a contact force between the two particles.
The effect of the contact force between the two closest points is equivalent to apply both a force and a torque on the center of masses of the particles. 
The force acting on the center of mass of particle $i$ due to particle $j$ is given by 
\begin{align}
	\textbf{f}_{ij}=-\dfrac{\partial \phi}{\partial \rv_i}=-\dfrac{\partial \phi}{\partial r}\dfrac{\rv}{r}=-\textbf{f}_{ji},
\end{align}
and the torque acting on particle $i$ due to particle $j$ is given by 
\begin{align}
	\textbf{t}_{ij}=-\uv_i\times\dfrac{\partial \phi}{\partial \uv_i}=\rv_i^c\times\textbf{f}_{ij}.
\end{align}
Here, $\rv_i$ and $\uv_i$ are the position of the center of mass and the orientation of particle $i$, respectively. The vector $\rv_i^c$ joins the center of mass of particle $i$ with the point of application of the force.

To calculate the minimum distance between two particles we calculate all possible corner-corner and corner-edge distances and select the minimum of all of them.

{\bf Verlet-type integration algorithm for Langevin dynamics.}
We calculate the trajectories following the integration scheme for Langevin dynamics presented in Ref.~\cite{GroenbechJensen2013} for isotropic particles.
The translational equations of motion for particle $i$ are given by
\begin{align}
\dot\rv_i=&\vel_i,\label{eq:LD1}\\
m\dot\vel_i=&\textbf{f}_i-\gamma\vel_i+\textbf{f}_i^\text{\,rand}.\label{eq:LD2}
\end{align}
Here $\rv_i$, $\vel_i$, and $\textbf{f}_i$ are the position, the velocity, and the total force (internal plus external) of particle $i$ (the overdot indicates time derivative), $m$ is the mass of one particle, $\gamma$ is the translational friction coefficient, and $\textbf{f}_i^\text{\,rand}$ is a delta-correlated Gaussian random force (described in detail in Appendix~\ref{A:BD}).
These equations are integrated with the following Verlet-type scheme~\cite{GroenbechJensen2013}
\begin{eqnarray}
	\rv_i(t+\Delta t)&=&\rv_i(t)+b\Delta t\vel_i(t)+\nonumber\\
	&&\dfrac{b\Delta t^2}{2m}\left[\textbf{f}_i(t)+\textbf{f}_i^\text{\,rand}(t+\Delta t)\right],\\
	\vel_i(t+\Delta t)&=&a\vel_i(t)+\dfrac{\Delta t}{2m}\left[a\textbf{f}_i(t)+\textbf{f}_i(t+\Delta t)\right]+\nonumber\\
	&&\dfrac{b\Delta t}{m}\textbf{f}_i^\text{\,rand}(t+\Delta t),
\end{eqnarray}
with the parameters $a=(1-\alpha)/(1+\alpha)$,  $b=1/(1+\alpha)$ and $\alpha=\gamma\Delta t/(2m)$.

The equations of motion for the angular degrees of freedom in our two-dimensional  system of rectangular particles following Langevin dynamics are given by 
\begin{eqnarray}
	\dot\varphi_i&=&\omega_i,\label{eq:r1}\\
	I\dot\omega_i&=&t_i-\gamma_r\omega_i+t_i^\text{rand}.\label{eq:r2}
\end{eqnarray}
Here $\varphi_i$, $\omega_i$, and $t_i$ are the azimuthal angle, the angular velocity, and the torque (internal and external) of particle $i$, $\gamma_r$ is the rotational friction coefficient, $t_i^{\text{rand}}$ is a random torque (see details in Appendix~\ref{A:BD}), and 
$I=m(L^2+D^2)/12$ is the moment of inertia around the axis normal to the particle that passes through the center of mass (the particles are assumed to have a homogeneous mass distribution).
The torques and the angular velocity point along the $z-$direction (normal to the particles).
Equations~\eqref{eq:r1} and~\eqref{eq:r2} have the same mathematical structure as Eqs.~\eqref{eq:LD1} and~\eqref{eq:LD2}. We therefore apply the same integration scheme as for the positional degrees of freedom, replacing the mass $m$ by the moment of inertia $I$ and the translational friction $\gamma$ by the rotational friction $\gamma_r$. For simplicity we set $\gamma_r=\gamma\sigma^2$. The value of the friction constants does not affect the equilibrium properties.

\section{Gay-Berne potential}\label{A:Gay-Berne}
We use the same implementation of the Gay-Berne potential as that in Ref.~\cite{Gay1981}.
The interaction potential between two particles is
\begin{eqnarray}
	\phi(\rv,\ui,\uk)&=&4\epsilon(\rv,\ui,\uk)\left[\left(\dfrac{\sigma_0}{r-\sigma(\rv,\ui,\uk)+\sigma_0}\right)^{12}\right.\nonumber\\
	&&\left.-\left(\dfrac{\sigma_0}{r-\sigma(\rv,\ui,\uk)+\sigma_0}\right)^{6}\right].
\end{eqnarray}
with $\rv$ the vector joining the centers of mass of the particles, $\ui$ and $\uk$ unit vectors along the long axes of the particles,
and the functions $\epsilon(\rv,\ui,\uk)$ and $\sigma(\rv,\ui,\uk)$ are given by
\begin{align}
    \epsilon(\rv,\ui,\uk)&=\epsilon_0\left(\xi_+\xi_-\right)^{-1/2}\left(\sigma^*(\chi')\right)^2,\\
    \sigma(\rv,\ui,\uk)&=\sigma_0\left(\sigma^*(\chi)\right)^{-1/2},
\end{align}
with
\begin{align}
	\chi&=\dfrac{l^2-1}{l^2+1},\\
	\chi'&=\dfrac{\sqrt{d}-1}{\sqrt{d}+1},\\
	\sigma^*(\xi)&=1-\dfrac{\xi}{2}\left[\dfrac{(\ru{+})^2}{\xi_+}+\dfrac{(\ru{-})^2}{\xi_-}\right],\\
	\xi_\pm&=1\pm\xi\ui\cdot\uk,\\
	\ru{\pm}&=\rv\cdot\ui\pm\rv\cdot\uk.
\end{align}
Here, $\xi$ takes the values $\chi$ or $\chi'$ and the parameters $\epsilon_0$ and $\sigma_0$ set the energy and the length scales.
We select a length-to-width ratio $l=3$, and set the energy ratio between the side-by-side and the tip-to-tip configurations to $d=5$. 

\section{Overdamped Brownian dynamics with orientational degree of freedom}\label{A:BD}
The equations of motion of particle $i$ are 
\begin{align}
	\gamma\vel_i&=\textbf{f}^\text{\,rand}_i(t)-\nabla_iu(\rv^N,\uv^N)-\nabla_i\Vext\left(\rv_i,\uv_i\right),\label{eq:motioncenter}\\
	\gamma_r\boldsymbol{\omega}_i&=\textbf{t}^\text{rand}_i(t)-\nablarot_i u\left(\rv^N,\uv^N\right)-\nablarot_i\Vext\left(\rv_i,\uv_i\right),\label{eq:motionangle}
\end{align}
Here, $\textbf{f}^\text{\,rand}_i$ and $\textbf{t}^\text{rand}_i$ are delta-correlated Gaussian random forces and torques, respectively, with zero means and variances
\begin{align}
    \left\langle\textbf{f}^\text{\,rand}_i(t)\textbf{f}^\text{\,rand}_k(t')\right\rangle&=2k_BT{\gamma}\identity\delta_{ik}\delta(t-t'),\label{eq:force-fluctuation}\\
    \left\langle\textbf{t}^\text{rand}_i(t)\textbf{t}^\text{rand}_k(t')\right\rangle&=2k_BT{\gamma_r}(\identity-\uv\uv)\delta_{ik}\delta(t-t').\label{eq:torque-fluctuation}
\end{align}
Here, the angles denote an average of the noise, $\identity$ is the identity matrix, $\uv\uv$ indicates the dyadic product, and $\delta_{ik}$ is the Kronecker delta.
The angular velocity is
\begin{equation}
\omegai=\uv_i\times\dot\uv_i.
\end{equation}
Using the vector triple product and the fact that $\uv_i\cdot\dot\uv_i=0$ due to $\uv_i\cdot\uv_i=1$, it follows directly that
\begin{equation}
	\dot\uv_i = \omegai\times\uv_i.
\end{equation}
Hence, the equations of motion can be integrated in time using the standard Euler algorithm via 
\begin{widetext}
\begin{align}
	\rv_i(t+\Delta t)&=\rv_i(t)+\dfrac{\Delta t}{\gamma}\left[-\nabla_iu(\rv^N,\uv^N)-\nabla_i\Vext\left(\rv_i,\uv_i\right)\right]+\boldsymbol{\eta}_i(t)\\
	\uv_i(t+\Delta t)&=\uv_i(t)+\dfrac{\Delta t}{\gamma_r}\left[-\nablarot_i u\left(\rv^N,\uv^N\right)-\nablarot_i\Vext\left(\rv_i,\uv_i\right)\right]\times\uv_i(t)+\boldsymbol\Gamma_i(t).
\end{align}
\end{widetext}
Here, $\boldsymbol{\eta}_i$ is a Gaussian random displacement with zero mean and standard deviation $\sqrt{{2k_BT\Delta t}/{\gamma}}$, and
\begin{equation}
	\boldsymbol\Gamma_i=\sqrt{\dfrac{2k_BT\Delta t}{\gamma_r}}(U^1_i\hat{\textbf{w}}^1_i+U^2_i\hat{\textbf{w}}^2_i)
\end{equation}
is a random rotation. Here, $U_i^1$ and $U_i^2$ are Gaussian random numbers with zero mean and unit width, and $\hat{\textbf{w}}^1_i=\ev_x\times\uv_i/|\ev_x\times\uv_i|$, $\hat{\textbf{w}}^2_i=\hat{\textbf{w}}^1_i\times\uv_i$, and $\textbf{u}_i$ are orthonormal to each other.
We renormalize $\uv_i$ after each time step such that it remains a unit vector.

We arbitrarily relate the rotational friction coefficient to the translational friction coefficient via $\gamma_r=\gamma\sigma^2$.
Also, we use a single translational friction coefficient $\gamma$ for displacements parallel and perpendicular to the main axis of the particle.
The values of the friction coefficients do not play any role in the equilibrium distribution functions.
The Euler integration time step is $\Delta t=10^{-4}\tau$ and we sample every $10^{-2}\tau$.
Although we have used here a simple Euler scheme to integrate the equations of motion, it would be useful to extend the recently developed adaptive Brownian dynamics~\cite{adaptiveBD} algorithm to systems with orientational degrees of freedom to speed up the simulations.

\end{document}